\documentclass[linenumbers]{aastex631}
\usepackage{makecell}

\nolinenumbers

\newcommand{\hunit}{$\rm km\,s^{-1}\,Mpc^{-1}$}

\begin{document}

\title{3D Localization of FRB\,20190425A for Its Potential Host Galaxy and Implications}

\author{Da-Chun Qiang}
\affiliation{Institute for Gravitational Wave Astronomy, Henan Academy of Sciences, Zhengzhou 450046, Henan, China}

\author{Zhiqiang You}
\affiliation{Institute for Gravitational Wave Astronomy, Henan Academy of Sciences, Zhengzhou 450046, Henan, China}

\author{Sheng Yang}\thanks{e--mail: sheng.yang@hnas.ac.cn}
\affiliation{Institute for Gravitational Wave Astronomy, Henan Academy of Sciences, Zhengzhou 450046, Henan, China}
\affiliation{INAF Osservatorio Astronomico di Padova, Vicolo dell’Osservatorio 5, I-35122 Padova, Italy}

\author{Zong-Hong Zhu}\thanks{e--mail: zhuzh@hnas.ac.cn}
\affiliation{Institute for Gravitational Wave Astronomy, Henan Academy of Sciences, Zhengzhou 450046, Henan, China}

\author{Ting-Wan Chen}
\affiliation{Graduate Institute of Astronomy, National Central University, 300 Jhongda Road, 32001 Jhongli, Taiwan}

\begin{abstract}

Fast radio bursts (FRBs) are high-energy, short-duration phenomena in radio astronomy. Identifying their host galaxies can provide insights into their mysterious origins. In this paper, we introduce a novel approach to identifying potential host galaxies in three-dimensional space. We use FRB\,20190425A and GW190425 as an example to illustrate our method. Recently, due to spatial and temporal proximity, the potential association of GW190425 with FRB\,20190425A has drawn attention, leading to the identification of a likely host galaxy, UGC\,10667, albeit without confirmed kilonova emissions. We search for the host galaxy of FRB\,20190425A with a full CHIME localization map. Regardless of the validity of the association between GW190425 and FRB\,20190425A, we identify an additional potential host galaxy (SDSS\,J171046.84+212732.9) from the updated GLADE galaxy catalog, supplementing the importance of exploring the new volume. We employed various methodologies to determine the most probable host galaxy of GW190424 and FRB\,20190425A, including a comparison of galaxy properties and constraints on their reported observation limits using various Kilonova models. Our analysis suggests that current observations do not definitively identify the true host galaxy. Additionally, the Kilonova models characterized by a gradual approach to their peak are contradicted by the observational upper limits of both galaxies. Although the absence of optical emission detection raises doubts, it does not definitively disprove the connection between GW and FRB.

\end{abstract}

\section{Introduction} \label{sec:intro}
Fast radio bursts (FRBs) are bright millisecond-duration radio bursts originating at cosmological distances \citep{Lorimer2007}. To date, nearly a thousand cases of FRB have been detected \citep{blinkverse}. These bursts are broadly categorized as repeating and apparently non-repeating in observations. The origin of FRBs is still unknown, although the repeating FRB\,20200428A has been traced back to a magnetar, i.e. SGR\,J1935+2154, in the Milky Way \citep{CHIMEFRB200428,Bochenek2020}. And there have been the X-ray emissions from this magnetar that coincide with FRB\,20200428A \citep[e.g.][]{2020ApJ...898L..29M,2021NatAs...5..378L}. Currently, many models have been proposed to explain the origin of FRBs. They can be derived into two kinds: repeating models and catastrophic models \citep[see][for an overview and references there in]{Platts2019,Xiao2021,Xiao2022,Zhang2023}. The former usually invokes a neutron star (NS) as the center engine of the FRB to explain repeating FRBs. The latter generally refers to compact star merger events such as binary neutron stars (BNS) or NS-BH (BH means black hole).

Identifying the host galaxy of FRB sources is crucial to understanding their origins. Thanks to the implementation of radio telescope arrays such as Australian Square Kilometre Array Pathfinder (ASKAP) radio telescope \citep{ASKAP_new} and Deep Synoptic Array \citep[DSA,][]{DSA-10}, the positioning accuracy of FRBs has now achieved sub-arcsecond precision. At present, about fifty FRB host galaxies have been found, mainly because of their precise localization, close proximity, or their repeating properties that allow radio telescopes to locate them accurately \citep[e.g.][]{DSA,2017Natur.541...58C,Bhandari2022,2020Natur.577..190M,2019Sci...365..565B,James2022}. Following the localization of FRBs, researchers can search for a potential host galaxy with the galaxy catalog by the PATH method \citep{path}. This method employs a Bayesian approach to calculate the probability that a galaxy is the host of an extragalactic transient source. It uses the sky coordinates and uncertainties of the transient source, along with the galaxy's flux and FRB radial offsets. However, this method only evaluates the probability that a galaxy is the host galaxy on a two-dimensional plane, and such results may not be accurate enough. The accuracy of the results will be improved if the distance of the source is taken into account. For an FRB source, its dispersion measure (DM)—the integral of the free electron number density along the propagation path—can serve as a distance proxy \citep[e.g.,][]{Deng2014ApJ,Yang2016ApJ,Ioka2003ApJ,Inoue2004MNRAS}. The DM is primarily contributed by our Galaxy, the intergalactic medium, and the host galaxy of the FRB source. However, the error in DM is too large for it to be a reliable third-dimensional indicator, as it is challenging to determine the contribution of the host galaxy and account for plasma density fluctuations in the intergalactic medium \citep{McQuinn2014ApJ,Ioka2003ApJ,Jaroszynski2019MNRAS}. It would be preferable to obtain the distance to the FRB source using other methods, such as the luminosity distance derived from gravitational waves (GWs).

Since the advanced Laser Interferometer Gravitational-Wave Observatory (LIGO) \citep{aLIGO2015} detected the first GW event, i.e. GW150914 \citep{2016PhRvL.116x1102A}, many GW events have been observed in the past decade, contributing significantly to our understanding of the universe \citep{Abbott2019O1O2, Abbott2021O3-1, Abbott2023O3-2}. The second generation GW detectors, e.g. LIGO, Virgo \citep{avirgo2015} and KAGRA \citep{kagra2012,kagra2013,kagra2018,kagra2019}, are designed to probe high frequency ($\sim$ 10--1000\,Hz) GW signals whose main astrophysical sources are compact binary coalescences (CBCs). Theoretical predictions suggest that when these systems include an NS, they can generate intense electromagnetic (EM) radiation. 

The first BNS merger was detected by LIGO as GW170817 \citep{Abbott2017PRLGW170817} on August 17, 2017, and hereafter associated counterparts were detected across the EM spectrum: a coincident short gamma-ray burst (GRB) 170817A \citep{Fermi_KN,INTEGRAL_KN}, an optical counterpart AT\,2017gfo \citep{sss17a,dlt17ck,lcogtkn,vista,master,decam} that resembled a kilonova (KN), as well as X-ray \citep{Haggard_17,Troja_17,Margutti_17} and radio \citep{Hallinan_17} counterparts that resembled off-axis jets. By combining GW170817 with various types of EM counterparts, this event has become a valuable tool for advancing research in astrophysics and cosmology. \citep[e.g.][]{Abbott2017GW170817H0,Abbott2018GW170817EoS}. However, we should note that the detection of the EM counterpart for GW170817 involved several fortuitous factors. Notably, this event occurred at a relatively close distance, only 40 Mpc, resulting in a well-enough localization by LIGO (approximately 20 square degrees at a 90 percent confidence level). It is worth noting that LIGO's detection range for BNS mergers can extend beyond this range, and in a broader spatial context, the precision of LIGO's localization may not be as high. This is exemplified in another BNS-generated GW event, GW190425 \citep{Abbott2020ApJGW190425}. There were no EM counterparts (such as a KN or a GRB) detected, due to its considerable distance ($d_{\rm L}=159 ^{+69}_{-71} \,{\rm Mpc}$) and poor GW localization (8,284 square degree at 90 percent confidence level). 

In addition to GW localization, if other signals coincide with GWs and are well-localized, they can serve as excellent EM triggers together with GWs. As such, FRBs could potentially serve as the EM counterpart of some GW events because they both related to compact stars. The co-production of GW and FRB is hypothesized to be explained by the `blitzar' mechanism \citep{Zhang2014ApJ}. When a BNS merger produces a supra-massive NS, and then collapses into a black hole, the closed magnetic field lines will break away from the NS and be ejected, its so-called `blitzar', this process could produce FRB \citep{Falcke2014A&A,Most2018ApJ}. This mechanism can also explain the internal X-ray plateaus observed in some long and short GRBs \citep{Zhang2014ApJ}. In a recent study, \cite{Moroianu2023NA} explored the potential correlation between GW190425 and an apparently non-repeating FRB event, FRB\,20190425A \citep{CHIME/FRBcatlog12021}. FRB\,20190425A have accurate localization of $\rm RA = 255.72\pm0.14^{\circ}$ and $\rm DEC = 21.52\pm0.18^{\circ}$ with $1\sigma$ uncertainty. According to the dispersion measure of FRB\,20190425A ($\mathbf{\rm DM_{obs} = 127.8 \, pc\,cm^{-3}}$), they give a maximum redshift of $z_{\rm max} < \sim 0.04$, corresponding to a luminosity distance of $d_{\rm L} < 200\,{\rm Mpc}$. Furthermore, the FRB signal reached Earth 2.5 hours later than the GW signal, a delay consistent with the expected lifetime of a supermassive NS (ranging from hundreds to thousands of seconds). Therefore, the GW and FRB were coincident in their localization, burst time, and distance. And \cite{Moroianu2023NA} found that a high correlation probability, i.e. the unrelated chance probability of a coincidence between FRB\,20190425A and GW190425 in the searched databases is estimated to be 0.0052 ($2.8\sigma$). However, \cite{Bhardwaj2023a}, \cite{Radice2024}, and \cite{Hernandez2024} argue that GW190425 and FRB\,20190425A cannot be associated. \cite{Bhardwaj2023a} even contends that remnants of the BNS merger cannot account for the formation of more than $1\%$ of the FRB sources. Despite this, the close localization, burst time, and distance between GW190425 and FRB\,20190425A suggest that intriguing connections may exist.

\begin{figure}
  \begin{center}
  \includegraphics[width=0.8\columnwidth]
  {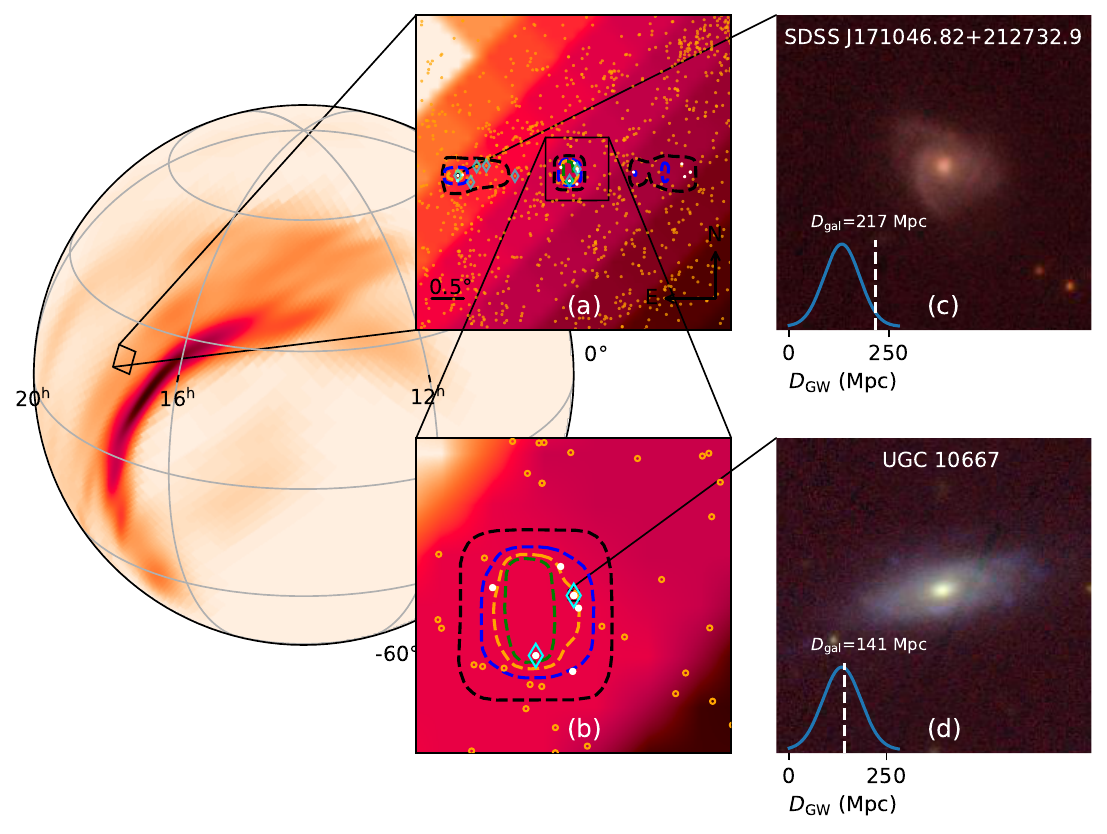}\\
  \end{center}
  \caption{\label{fig:selection}\textbf{The positions of the selected galaxies, along with GW190425, and FRB\,20190425A in a two-dimensional spherical coordinate system.} The panel (a) provides a close-up of the GLADE galaxies (open orange circles) that meet our criteria. The colored boxes indicate different confidence level regions of FRB\,20190425A, and the shaded region represents the probability of GW190425. The panel (b) shows a zoom-in view containing UGC\,10667. The galaxies in white point and cyan diamonds are the most probable galaxies selected by this study and \cite{Panther2023}, respectively. 
  The panel (c) and (d) presents archived PS1 colored images of SDSS\,J171046.82+212732.9 and UGC\,10667, respectively. The blue lines denote the probability distribution function of $d_{\rm L}$ estimated by GW190425. The white dashed lines are the $d_{\rm L}$ of galaxies, which are calculated with spectroscopic redshifts and a flat $\Lambda$CDM cosmology.}
\end{figure}

The localization accuracy of FRB\,20190425A was significantly better than that of GW190425. If their association is valid, the improved FRB localization, along with the precise GW luminosity distance, would substantially enhance the chances of mapping EM emissions.
This was investigated in \cite{Moroianu2023NA}: according to the $1\sigma$ uncertainty of CHIME localization and the $1\sigma$ upper limit of the LIGO luminosity distance, they found only one galaxy, i.e. UGC\,10667, as a potential host for both GW190425 and FRB\,20190425A.
\cite{Panther2023} search for potential host galaxies of FRB\,20190425A using the NASA Extragalactic Database, and they confirmed that UGC\,10667 is the most probable host galaxy.
However, we note that the localization map of FRB\,20190425A adopted in \cite{Panther2023} is not consistent with CHIME\footnote{https://www.chime-frb.ca/catalog/FRB20190425A}, which may potentially lead to mis-selection or omission of galaxies for subsequent EM analysis.
\cite{Bhardwaj2023b} searched for the host galaxy of the FRB\,20190425A in the more complete DESI catalog with CHIME baseband data, and they also confirmed that UGC\,10667 is the only galaxy within the $2\sigma$ localization region of the FRB. However, DESI may also be incomplete. 

If the association between GW190425 and FRB\,20190425A is confirmed, the host galaxy of GW190425 can be identified, which will aid in detecting KN radiation. This rapidly drew widespread attention from the astronomical community, and a KN search campaign was immediately carried out \citep[e.g.][]{Coughlin2019,Boersma2021,Paterson2021}. For instance, \cite{Smartt2024} conducted an optical follow-up search using ATLAS \citep{atlas2018} and Pan-STARRS \citep{panstarrs2016}. They covered 24.9 percent and 41.2 percent of GW190425 localization within 6.0 hours after the GW detection respectively, as well as imaging the galaxy UGC\,10667 3.5 hours after the FRB\,20190425A. 
Despite not detecting any optical counterparts, they compared their observational limits with various KN models and found that certain models could meet these limits, making it impossible to conclude whether the galaxy is the host based solely on the absence of KN radiation.

Moreover, combined with host galaxy redshift information, a GW event can be a bright standard siren to constrain the Hubble constant.
Current data from the early universe, as indicated by the Cosmic Microwave Background (CMB), report the latest Hubble constant $H_0$ of $67.4 \pm 0.5\,$\hunit\citep{Planck2018}. 
In contrast, a recent result from Type Ia supernova (SN Ia) observations, probing the late universe, suggests $H_0=73.04 \pm 1.04\,$\hunit\citep{sh0es}.
The significant disparity between these two measurements, exceeding $4\sigma$, goes beyond a level of chance, leading to what is known as the ``Hubble tension" \citep[e.g.][]{Riess2020h0tension,Perivolaropoulos2022,Mazo2022,Kamionkowski2023}.
Given this discrepancy, a third independent observation to measure $H_0$ becomes crucial in resolving this conundrum.
GWs are independent of early and late universe observations and are expected to help alleviate the Hubble tension.

In this paper, we assume an association between GW190425 and FRB\,20190425A, and use it as an example to show how we search and analyze potential host galaxies.
We believe that these studies need to take into account the completeness of the galaxy sample. Therefore, we decided to search for the host galaxy in the updated GLADE+ galaxy catalog \citep{GLADE+2022}. 
Referring to Figure 4 in \cite{GLADE+2022}, it is clear that the completeness of the GLADE+ catalog exceeds 90\% at the distance of GW190425.
In the meanwhile, the host galaxy of the KN (AT 2017gfo), NGC4993, is close to the peak of the B-band absolute magnitudes in the GLADE+ galaxy catalog.
Consequently, the absence of fainter galaxies in the GLADE+ catalog is improbable to serve as host galaxies of KN. However, we cannot entirely rule out the possibility that the origin of FRB\,20190425A is not within the GLADE+ galaxy catalog.
Additionally, beyond the $P_{\rm \, PATH}$ method employed by \cite{Panther2023}, we introduce a novel $P_{tmd}$ approach for galaxy prioritization, which balances additional constraints from the GW channel.
Using the localization maps of FRB\,20190425A from CHIME and GW190425 from LIGO, we identified several potential host galaxies. This is achieved by calculating the probabilities $P_{\rm \, PATH}$ and $P_{tmd}$, and fitting these galaxies to known FRB host galaxy models. Additionally, we fit kilonova (KN) models to their reported observational limits to further constrain the host galaxy candidates.
Finally, we report the Hubble constant constrained by combining the $H_0$ posterior from GW170817 and GW190425 with the potential host galaxies.

\begin{table}
\renewcommand{\arraystretch}{1.5}  
\setlength{\tabcolsep}{2pt}
\begin{center}
\begin{tabular}{ccccccccccc}\hline\hline
$\rm Identifier$ & $\rm RA(J2000)$ & $\rm DEC(J2000)$& $\rm Redshift$& $M_{\rm B}$& $p_{\rm  GW}*10^{-6}$ & $ p_{\rm FRB}$ & $ p_{d}$ & $P_{tmd}$& $P_{\rm \, PATH}$  \\ \hline
$\textbf{\rm UGC\,10667} $&$\textbf{\rm 255.66248}  $&$\textbf{\rm 21.57674} $&$\textbf{\rm 0.03122} $&$\textbf{\rm -20.44} $&$\textbf{\rm 3.80}$&$\textbf{\rm 0.36195}$&$\textbf{\rm 0.00810}$&$\textbf{\rm 0.58757} $&$\textbf{\rm 0.17108} $ \\
$\textbf{\rm SDSS\,J171046.82+212732.9} $&$\textbf{\rm 257.69510} $&$\textbf{\rm 21.45914}  $&$\textbf{\rm 0.04734} $&$\textbf{\rm -20.73} $&$\textbf{\rm 3.14}$&$\textbf{\rm 0.47314}$&$\textbf{\rm 0.00145}$&$\textbf{\rm 0.14804} $&$\textbf{\rm 0.39793}$ \\
$\rm SDSS\,J170310.06+212309.7$&$255.79194$&$21.38603$&$0.04789$&$-18.04$&$3.80$&$0.65994$&$0.00178$&$0.02570$&$0.01230$ & \\
$\rm SDSS\,J171048.21+212253.8$&$257.70092$&$21.38162$&$0.05026$&$-20.08$&$3.14$&$0.27481$&$0.00078$&$0.02555$&$0.06907$  \\
$\rm SDSS\,J165518.73+212637.8$&$253.82806$&$21.44384$&$0.03089$&$-17.94$&$4.32$&$0.03115$&$0.00709$&$0.00502$&$0.00052$  \\
$\rm SDSS\,J170239.90+212008.9$&$255.66629$&$21.33583$&$0.02556$&$-16.83$&$3.80$&$0.07786$&$0.00730$&$0.00408$&$0.00041$  \\
$\rm SDSS\,J165530.04+213843.2$  &$253.87520$&$21.64535$&$	0.04871$&$	-19.53$&$	4.32$&$	0.02032$&$	0.00205$&$	0.00408$&$	0.00225$\\
 \hline
$\rm SDSS\,J170249.74+214008.3$ &$255.70726$&$21.66898$&$0.03527$&$-18.68$&$3.80$&$0.27999$&$0.00698$&$0.07735$&$0.01072$\\
$\rm SDSS\,J170345.46+213605.3 $&$255.93946$&	$21.60147$&	$0.04428$&	$-18.77$&	$3.80$&	$0.21238$&	$0.00294$&	$0.02675$&	$0.00901$\\
$\rm SDSS\,J165455.49+213054.2$  & $253.73125	$&$21.51508$&$	0.04660$&$	-21.38	$&$4.32$&$	0.01524	$&$0.00277$&$	0.02285	$&$0.06914$\\
$\rm SDSS\,J170235.08+213211.5$  & $	255.64620$&$	21.53655	$&$0.05232$&$	-18.37$&$	3.80$&$	0.36261$&$	0.00078$&$	0.00844	$&$0.00971$\\
$\rm SDSS\,J171028.78+212816.4$  & $	257.61995$&$	21.47125	$&$0.05043	$&$-18.67$&$	3.14$&$	0.34436$&$	0.00075$&$	0.00838$&$	0.01301$\\
$\rm SDSS\,J165847.87+213119.5$  & $	254.69946$&$	21.52210$&$	0.04813$&$	-18.54$&$	4.26$&$	0.08809$&$	0.00216$&$	0.00743$&$	0.00286$\\
$\rm SDSS\,J171011.06+213858.5$  & $	257.54611$&$	21.64959$&$	0.01479$&$	-19.11$&$	3.14$&$	0.02123$&$	0.00315$&$	0.00326$&$	0.00137 $\\
\hline
\end{tabular}
\end{center}
  \caption{\label{tab:gals}\textbf{The Candidate host galaxies of GW190425 and FRB\,20190425A.} The sum of $P_{tmd}$ of these 14 galaxies is $ \sim 2 \sigma$. all galaxies are located in $99\%$ localization confidence of the CHIME localization with redshift $z<0.06$. The results presented in the top half of the table correspond to the golden sample, whereas the bottom half represents the bronze sample. Given are object name, RA and DEC, redshift of galaxy (spectral redshift for golden sample, and photometric redshift for bronze sample), B-band absolute magnitude $M_{\rm B}$, prior on GW probability $p_{\rm GW}$ and FRB probability ($p_{\rm FRB}=1-{\rm CL_{CHIME}}$ where $\rm CL_{CHIME}$ is the localization confidence level), prior of distance of galaxy $p_d$, posterior probability $P_{tmd}$ as defined by Eq.\ref{eq:ptmd}, and posterior probability $P_{\rm \, PATH}$ as defined in \cite{Panther2023}.}
\end{table}

\section{Identification of Host Galaxy\label{sec:idhost}}

\subsection{Probbility of Host Galaxy\label{subsec:ptmd}}
To assess the likelihood of a galaxy being the true host of an FRB, researchers frequently employ the `PATH' methodology developed by \cite{path}. This approach is grounded in a Bayesian framework, amalgamating the localization probability of an FRB inferred from radio observations with prior assumptions regarding the magnitudes of FRB host galaxies, radial offsets, and the probability that the actual host remains unobserved. For example, \cite{Panther2023} employed this method in the selection process, identifying galaxy UGC\,10667 as having the highest probability of being the host of FRB\,20190425A, while opting to overlook the potential radial offsets for each considered host galaxy, because the FRB localization uncertainty is much larger than the angular extent of the candidate host galaxies. Meanwhile, the process of prioritizing target galaxies associated with GW detections closely resembles that of FRBs, emphasizing factors such as GW probability and luminosity distributions, as detailed in \cite{yang2019}, notably in their section 3.1.

In our study, assuming the validity of the GW-FRB association, we integrate information from both signals. Specifically, we leverage the luminosity distance derived from the GW signals and the localization data from the FRB side. This comprehensive amalgamation of data enables us to establish a new criterion incorporating distance constraints from GW signals, enhancing the precision in the identification of potential host galaxies.
Our approach involves synthesizing diverse data points from GW and FRB signals, culminating in a new criterion that integrates distance constraints from GW signals to refine the identification process of likely host galaxies.

We establish a novel criterion termed `Trigger-Magnitude-Distance' (TMD) to depict the posterior probability of candidate host i. This criterion is defined as follows:

\begin{equation}\label{eq:ptmd}
P_{tmd}^{i} = \frac{p_{\,t}^{\,i}\, p_{\,m}^{\,i} \,p_{\,d}^{\,i}}{\sum_{j} \, {p_{\,t}^{\,j} \, p_{\,m}^{\,j} \, p_{\,d}^{\,j}}},
\end{equation}
where $p_{t}$, $p_{m}$, and $p_{d}$ represent priors on the trigger probability, magnitude, and distance of the galaxy, respectively. In this paper, we assume that the GW and FRB probability are equally weighted, thus:

\begin{equation}
p_{t}^{i} = p_{\rm {\,GW}}^{\,i} \, p_{\rm {\,FRB}}^{\,i}.
\end{equation}
Here, $p_{\rm \,FRB}=1-{\rm CL_{CHIME}}$, where $\rm CL_{CHIME}$ is the localization
confidence level at the position of the candidate galaxy. $p_{\rm GW}$ represents the probability of a galaxy being the host of a GW signal.

For the term $p_{m}$, we followed \cite{yang2019}, that is, mapping selected GLADE+ galaxies into celestial sphere with the HEALPIX tool \citep{healpix}, weighted by galaxy luminosity \footnote{We assume that the mass distribution follows the B-band luminosity.}, and smoothed with a Gaussian corresponding to each galaxy's reported radius. This yields a luminosity distribution map, and $p_{m}$ is subsequently derived at the reported position of the particular galaxy.

The term $p_{d}$ considers the potential association of a galaxy with GW triggers, taking into account distance information. Specifically, the LIGO-VIRGO-KAGRA collaboration releases the 3D localization for an individual GW candidate, presenting the distance likelihood along with the 2D probability distribution in each direction. Consequently, $p_{d}$ is depicted as the probability density function of the candidate galaxy concerning the GW distance estimation in the same direction.

\begin{figure}
\begin{center}
  \includegraphics[width=0.8\columnwidth]{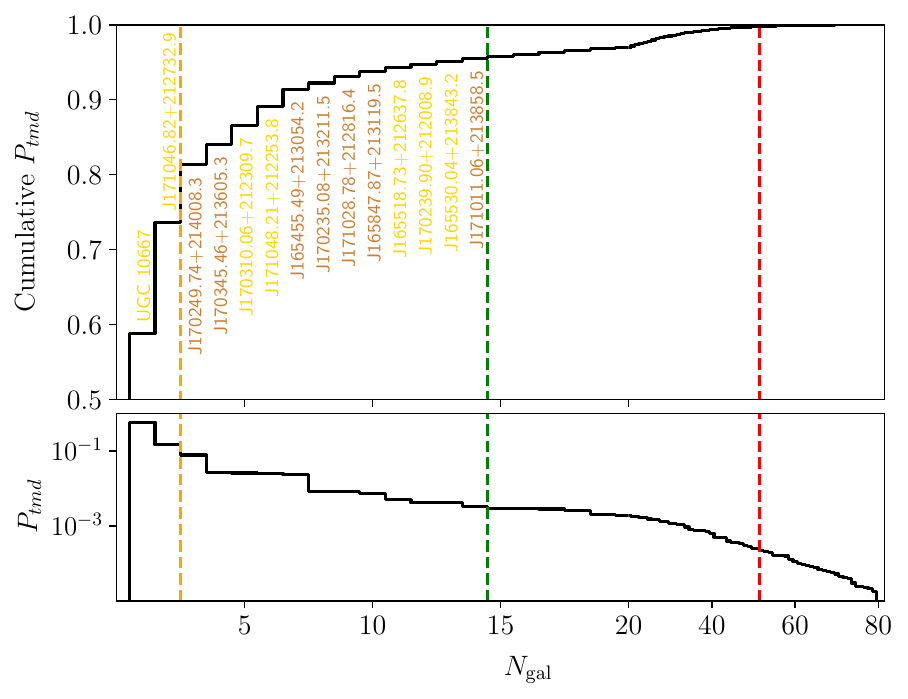}\\
\end{center}
\caption{\label{fig:ptmd}\textbf{The probability of all potential host galaxies.} The upper panel displays the cumulative $P_{tmd}$ of these galaxies with a black step line, while the lower panel shows their individual $P_{tmd}$. 
The orange, green and red dashed lines correspond to cumulative $P_{tmd}$ values of 0.73561, 0.95451 (approximately $2\sigma$) and 0.99749 (approximately $3\sigma$), respectively. 
The names of these top 14 galaxies are displayed in golden and bronze colors, representing their classification into the golden sample or bronze sample.}
\end{figure}

\begin{figure}
\begin{center}
  \includegraphics[width=0.6
\columnwidth]{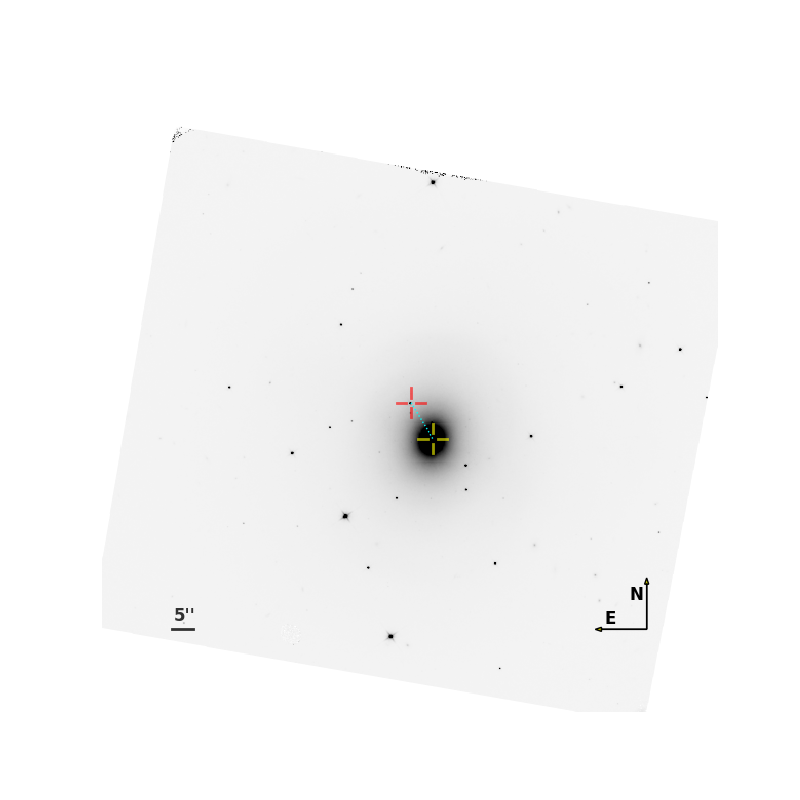}\\
  \end{center}
  \caption{\textbf{The archival HST image showing the offset from the kilonova to the host galaxy NGC4993.} }
\label{fig:kn2host}
\end{figure}

\subsection{Serching for Potential Hosts \label{subsec:sehost}}
By employing the intersecting localization regions of GW190425 and FRB\,20190425A, we conducted a search within the updated GLADE+ galaxy catalog \citep{GLADE+2022} for galaxies with redshift $z<0.06$\footnote{The dispersion measure of FRB\,20190425A is $\rm DM_{obs} = 127.8 \, pc\,cm^{-3}$, minus the contribution from the intergalactic medium in the Milky Way ($\rm DM_{MW,ne2001}=79.4\, pc\,cm^{-3}$), we get a maximum redshift of $z_{\rm max}=0.058$ \citep[assume the flat $\Lambda$CDM with parameters from][] {Planck2018}.}. We found more than one thousand candidate galaxies, represented by orange open circles in Figure~\ref{fig:selection}.
We calculated and normalized the probability ($P_{tmd}$) for each of them and found 81 galaxies with reasonable $P_{tmd}$. 
The probability density function (PDF) and the cumulative distribution function (CDF) of their $P_{tmd}$ values are illustrated in Figure~\ref{fig:ptmd}.
A comprehensive map, denoted as $P_{tmd}$, is subsequently normalized to represent the probability of a galaxy matching various trigger information and being sufficiently luminous to host a kilonova (KN). 
In Figure\,\ref{fig:ptmd}, both the collective and individual $P_{tmd}$ values for all potential host galaxies are displayed. 
The black step line represents the cumulative $P_{tmd}$ of these potential hosts.
The orange, green, and red dashed lines correspond to cumulative $P_{tmd}$ values of 0.73561, 0.95451 ( $\sim 2\sigma$), and 0.99749 ( $\sim 3\sigma$), respectively. 
Observing the cumulative contribution of the top 14 galaxies, it is evident that it encompasses approximately $2\sigma$. 
In contrast, the number of galaxies between $2\sigma$ and $3\sigma$ is disproportionately high, with considerably lower values of $P_{tmd}$. 
Hence, the top 14 galaxies are adequate for this study. 
Their names are presented in both golden and bronze colors, indicating their classification into the respective golden sample or bronze sample (see Sec.\ref{subsec:hostan}).
According to our method, we identified an additional galaxy (SDSS\,J171046.84+212732.9) as being a possible host galaxy for GW190425 and FRB\,20190425A. 

As shown, the significant contribution originated from the top 14 galaxies (listed in Table~\ref{tab:gals}) with the highest $P_{tmd}$. This yielded a probability of $0.95451$ ($\sim 2\sigma$), prompting us to retain these galaxies for further investigations.
In the two panels in the middle of Figure~\ref{fig:selection}, these selected 14 galaxies are localized as white points. For comparison, the selected galaxies in \cite{Panther2023} are overlaid with open cyan diamonds.
The dashed overlaid boxes and shaded colors indicate the various confidence level regions of FRB\,20190425A and GW190425.

Indeed, among these 14 galaxies, UGC\,10667 and SDSS\,J171046.84+212732.9 collectively account for over 70 percent of the $P_{tmd}$ probability, making them more likely to be the host galaxy. 
Following the approach outlined in \cite{Panther2023}, we computed the $P_{\rm \, PATH}$ probability for these galaxies as well. Interestingly, although UGC\,10667 has a higher value of $P_{tmd}$ compared to SDSS\,J171046.84 + 212732.9, it has a relatively lower value of $P_{\rm \, PATH}$. Therefore, both galaxies continue to be strong contenders.

For both galaxies, we downloaded the $ugriz$ frames from the Pan-STARRS (PS1) archive and stacked them into colored images using the `Lupton' method \citep{Lupton2004}.
These images are presented in two panels on the left side of Figure~\ref{fig:selection}. We explored the morphological features of these two galaxies, and found that UGC\,10667 and SDSS\,J171046.84+212732.9 are both classified as spiral galaxies.
Within the two panels, the blue lines represent the PDF of the luminosity distance ($d_{\rm L}$) estimated by GW190425, and the white dashed lines indicate the $d_{\rm L}$ values of galaxies calculated using spectroscopic redshifts and a standard $\Lambda$CDM cosmology.

Additionally, we consider the radial offset between the potential KN and the core of the galaxy. We quantified the spatial separation between AT\,2017gfo and NGC4993 using the archived Hubble Space Telescope (HST) image \footnote{We retrieved observations of AT\,2017gfo taken by the Wide Field Camera 3 (WFC3) from the MAST archive (PI: A. Levan). The HST proposal ID is 14804, and the image data is from \url{https://archive.stsci.edu/proposal_search.php?id=14804&mission=hst}. We used astrodrizzle to optimally regrid the image and SExtractor \citep{sextractor} to identify point-like or the center of extended sources.} ,and assumed that this distance is universally applicable to all KN locations.
Specifically, we observe AT\,2017gfo having an angular offset of $\sim 10.4''$ from the galactic core (see Figure~\ref{fig:kn2host}), corresponding to a linear diameter of $\sim 2.5 \, {\rm kpc}$\footnote{This is a projected distance, representing a lower limit on the actual radial distance of the KN from the galaxy nucleus.} at a luminosity distance of $40.7 \,{\rm Mpc}$ \citep{ngc4993dist}. At the distance of GW190425 (that is, 159 Mpc), this offset translates to approximately ${2.66''}$, equivalent to approximately 1 pixel in the HEALPIX framework with $N_{side}$=1,024. Therefore, when calculating the $P_{tmd}$ of a galaxy, we compute the value of the pixel corresponding to the direction of the galaxy, as well as the values of the surrounding 9 pixels, and then take the average.

\begin{figure}
\begin{center}
\includegraphics[width=0.95\columnwidth]{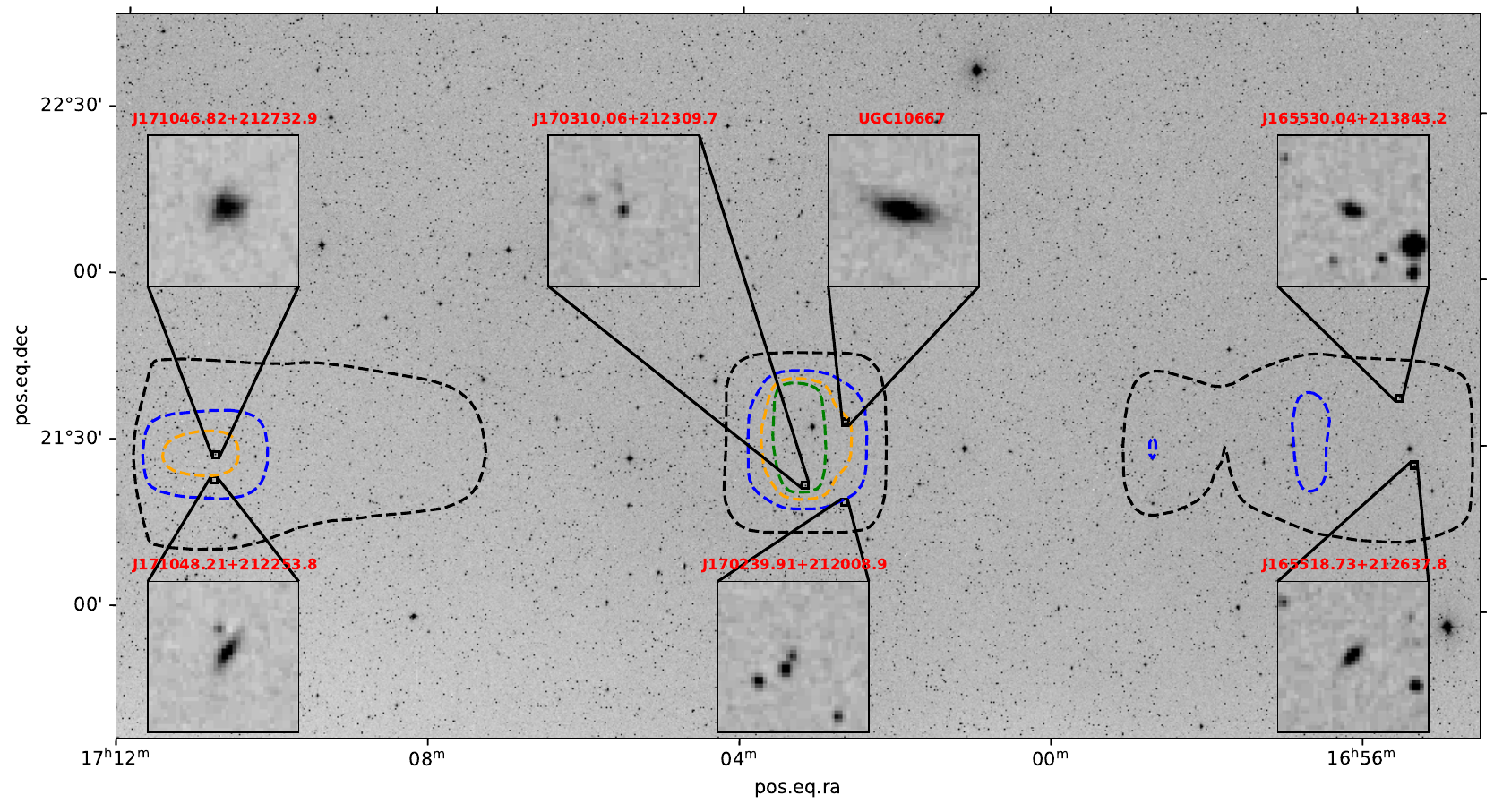}\\
\end{center}
\caption{\label{fig:gals}\textbf{The localization and morphology of seven galaxies in golden sample.} We exclude seven other candidates that do not have reliable redshift information. The dashed contours and colors indicate different confidence level regions of FRB\,20190425A.} 
\end{figure}

\begin{table}
\centering
\renewcommand{\arraystretch}{1.5}  
\setlength{\tabcolsep}{8pt}
    \begin{tabular}{cccccc}\hline\hline
    $\rm Identifier$ &
    \makecell[c]{\hspace{-1.2cm}$\rm Stellar\ mass$\\\hspace{-1.2cm} $\rm log(M/{M}_\odot)$ } 
    & \makecell[c]{\hspace{-1.2cm} $\rm SFR $\\ \hspace{-1.2cm}$\rm ({M}_\odot/yr)$ }&\makecell[c]{\hspace{-1.2cm}$\rm Stellar\ metallicity $\\ \hspace{-1.2cm}$\rm (Z/{Z}_\odot)$} & \makecell[c]{\hspace{-1.2cm}$\rm Light\!-\!weighted\ age$\\ \hspace{-1.2cm}$\rm \log(yr)$} &\makecell[c]{\hspace{-1.2cm}$\rm  Log \ sSFR $\\ \hspace{-1.2cm}$\rm (yr^{-1})$}\\ \hline
    ${\rm UGC\,10667}$ & $10.26^{+0.10}_{-0.09}$ &  $1.61^{+3.67}_{-1.03}$ & $0.72^{+0.74}_{-0.50}$ & $9.35^{+0.23}_{-0.29}$ & $-10.07^{+0.55}_{-0.45}$ \\
    ${\rm SDSS\,J171046.82+212732.9}$&  $10.55^{+0.10}_{-0.10}$ &  $1.39^{+3.77}_{-0.97}$ &   $0.77^{+0.69}_{-0.51}$ &   $9.48^{+0.21}_{-0.27}$ &   $-10.42^{+0.54}_{-0.50}$ \\
    $ \rm SDSS\,J170310.06+212309.7$&  $9.22^{+0.12}_{-0.13}$ &   $0.04^{+0.14}_{-0.04}$ &   $0.75^{+0.74}_{-0.48}$ &   $9.38^{+0.24}_{-0.27}$ &   $-10.57^{+0.55}_{-0.75}$\\ 
    $\rm SDSS\,J171048.21+212253.8$ &   $10.51^{+0.10}_{-0.11}$ &   $1.36^{+5.68}_{-1.08}$ &   $0.99^{+0.65}_{-0.62}$ &   $9.48^{+0.24}_{-0.33}$ &   $-10.37^{+0.64}_{-0.65}$\\
    $\rm SDSS\,J165518.73+212637.8$&   $9.68^{+0.10}_{-0.09}$ &    $0.44^{+0.94}_{-0.28}$ &   $1.14^{+0.58}_{-0.66}$ &   $9.40^{+0.21}_{-0.33}$ &   $-10.07^{+0.55}_{-0.40}$ \\
    $\rm SDSS\,J170239.90+212008.9$ &   $7.55^{+0.40}_{-0.08}$ &   $0.31^{+0.05}_{-0.06}$ &   $0.26^{+0.29}_{-0.20}$ &   $7.78^{+0.91}_{-0.19}$ &   $-8.08^{+0.10}_{-0.45}$\\
    $\rm SDSS\,J165530.04+213843.2$&   $10.17^{+0.10}_{-0.12}$ &   $0.39^{+1.30}_{-0.30}$ &   $0.78^{+0.72}_{-0.50}$ &   $9.49^{+0.23}_{-0.26}$ &   $-10.57^{+0.60}_{-0.60}$\\   
    \hline
    \end{tabular}
\caption{\label{tab:sdss} \textbf{Galaxy properties estimated using $\rm \textbf{MAGPHYS}$ based on photometric and spectral data from SDSS DR17.} The errors of all parameters are in $1\sigma$ uncertainty.}
\end{table}

\begin{figure}
\begin{center}
  \includegraphics[width=0.8\columnwidth]{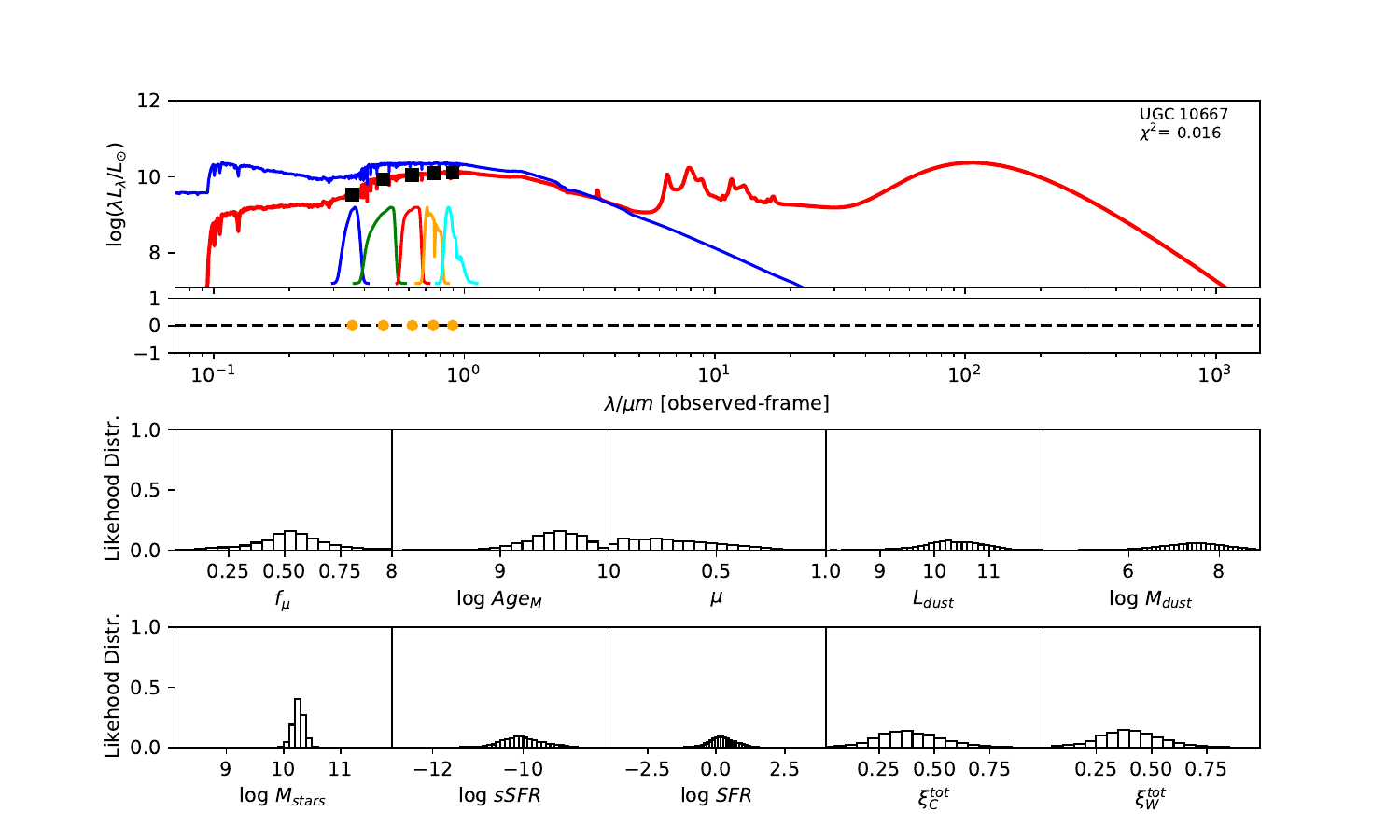}\\
  \end{center}
  \caption{\textbf{The figure displays the best-fit and likelihood distribution of a sample of stellar evolution and population synthesis models fitted to the photometric data of UGC\,10667.} The upper panel shows the best-fit model from $\rm \textbf{MAGPHYS}$ (red line) on SDSS DR17 modelMag data (filled black squares) of UGC\,10667. The blue line is the predicted unattenuated stellar population SED for the best-fit model. The colored unimodal lines represent the transmission curves of the u, g, r, i, and z filters in the SDSS system. The lower panel shows the residual between the observed and model photometry. The bottom two rows are full likelihood distributions of galaxy properties (The meanings of these galaxy properties can be found in \cite{da_Cunha2008})}.
\label{fig:galfit1}
\end{figure}

\begin{figure}
\begin{center}
   \includegraphics[width=0.8\columnwidth]{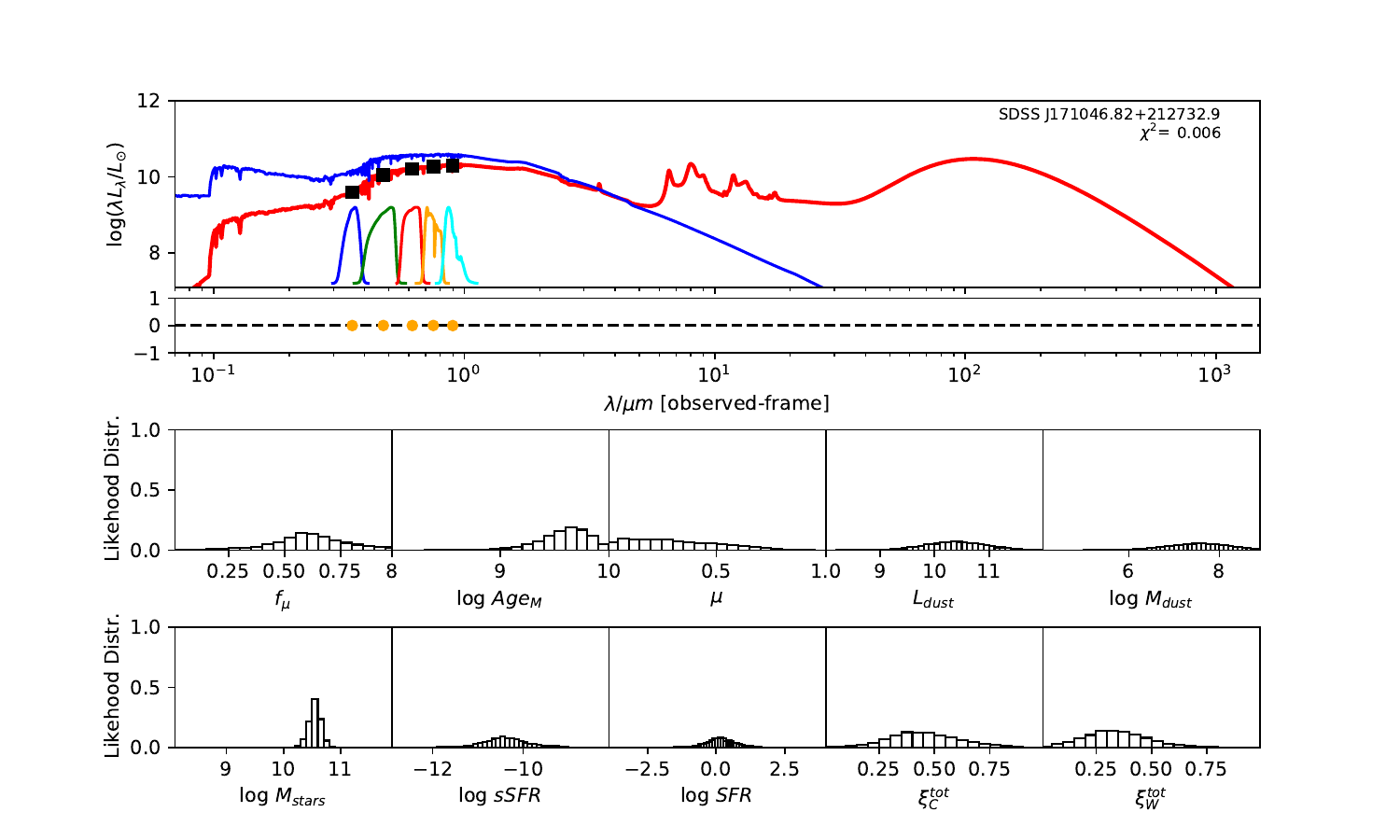}\\
  \end{center}
  \caption{\textbf{Similar to Figure~\ref{fig:galfit1}, this figure depicts the corresponding information for SDSS\,J171046.82+212732.9.} } 
\label{fig:galfit2}

\end{figure}

\subsection{Host Galaxy Analysis\label{subsec:hostan}}
To further confirm the host galaxy, we need spectral data of the galaxy. Therefore, we conducted cross-matching within the Sloan Digital Sky Survey (SDSS) DR17 for the 14 GLADE galaxies. Of these, 7 galaxies (including both UGC\,10667 and SDSS\,J171046.84+212732.9) were observed spectroscopically by SDSS, designated as our golden sample (its location and morphology are shown in Figure~\ref{fig:gals}), while the remaining 7 were classified as the bronze sample.

To further pinpoint the actual host galaxy of GW190425 and FRB\,20190425A, we performed a comparative analysis of the properties of the golden sample galaxies. We adopted model magnitudes (modelMag) from SDSS DR17 \citep{sdssdr17} for galaxies, including multicolors $ugriz$. After correcting for Milky Way extinction \citep{Schlafly2011}, these magnitudes were used to fit spectral energy distribution (SED) models to estimate stellar mass. The $\rm \textbf{MAGPHYS}$ stellar population model program \citep{da_Cunha2008} was employed for this purpose. The program utilizes a library of stellar evolution and population synthesis models from \cite{Bruzual2003} and performs fitting procedures to ascertain the luminosity of the host stellar population. $\rm \textbf{MAGPHYS}$ computes the total stellar mass, stellar metallicity compared to the solar abundance, and star formation rate (SFR), and also computes the light-weighted age of the host stellar population in the $r$-band. This represents an average age considering all the stars in the galaxy. The specific star formation rate (sSFR) is defined as the SFR divided by stellar mass.
$\rm \textbf{MAGPHYS}$ further computes the probability density function across various model values and determines the median, along with the confidence interval corresponding to the 16th to 84th percentile range (equivalent to the $1\sigma$ range, assuming a Gaussian distribution). We consider the median as the best estimate of the stellar mass, which is reported in Table~\ref{tab:sdss}. 

For UGC\,10667, the median stellar mass is $1.8\times10^{10}$\,M$_\odot$, with a range of $1\sigma$ that spans from $1.5\times10^{10}$ to $2.3\times10^{10}$\,M$_\odot$. The results are consistent with \cite{Bhardwaj2023b} because we use a similar code and prior. The other galaxy, SDSS\,J171046.82+212732.9, exhibits a slightly larger stellar mass of $3.6\times10^{10}$\,M$_\odot$, with a range from $2.8\times10^{10}$ to $4.5\times10^{10}$\,M$_\odot$. 
Most of our identified galaxies exhibit a stellar mass ranging from $1.7\times10^{9}$ to $3.6\times10^{10}$\,M$_\odot$, with stellar population ages around 2-3\,Gyr. However, one outlier, SDSS\,J170239.90+212008.9, stands out as a blue star-forming galaxy based on SDSS colors and the detection of strong emission lines in the spectrum. This aligns with our $\rm \textbf{MAGPHYS}$ fitting results, indicating a high specific star formation rate (sSFR) of approximately $8.4$\,Gyr$^{-1}$, a young stellar population age of around $60$\, Myr, and a low stellar metallicity of about $0.3$\,$\rm Z_\odot$. Specifically, we presented the SED model fitting results of UGC\,10667 and SDSS\,J171046.82+212732.9 by $\rm \textbf{MAGPHYS}$ in Figure~\ref{fig:galfit1} and \ref{fig:galfit2}. As shown, the intrinsic properties of galaxies UGC\,10667 and SDSS\,J171046.84+212732.9 are similar, which is consistent with their similar morphological characteristics.

\begin{figure}
\begin{center}
  \includegraphics[width=0.9\columnwidth]{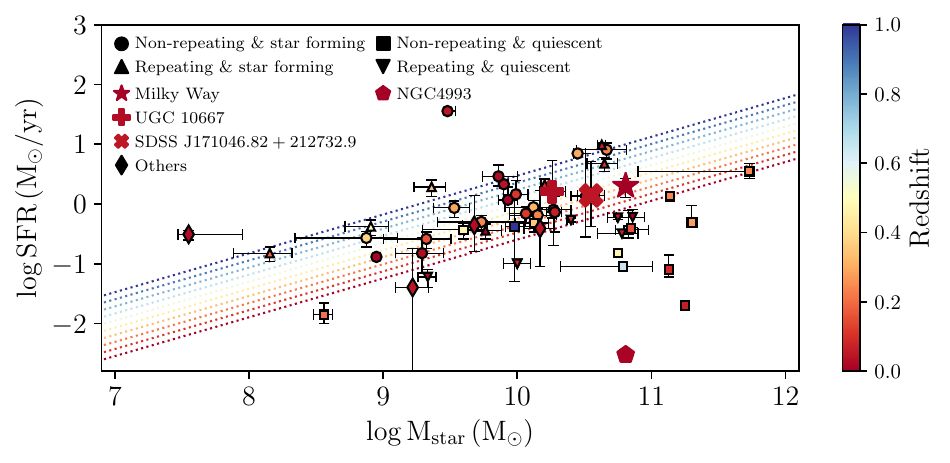}\\ 
\end{center}
\caption{\label{fig:sfr_vs_mstar}\textbf{Star formation rate and stellar mass distribution of the host galaxies of 32 apparently non-repeating and 14 repeating FRBs together with the Milky Way and golden sample galaxies.} The different colors of dotted lines and markers represent the redshift. The dotted lines represent the boundary that separates star forming and quiescent galaxies taken from the PRIMUS survey \citep{sfr_vs_mstar}. Circles and triangles represent non-repeating and repeating FRBs hosted by star-forming galaxies, respectively, while squares and inverted triangles denote the same for quiescent galaxies. The Milky Way, UGC\,10667, SDSS\,J171046.82+212732.9, NGC4993, and other golden sample galaxies are shown as a star, a plus sign, a cross, a pentagon, and diamonds, respectively. Data on the star formation rate and stellar mass of host galaxies for 32 apparently non-repeating and 14 repeating FRB can be found in \cite{Gordon2023,Bhardwaj2021a,Bhardwaj2021b,Bhandari2021,Bhandari2022,Ravi2019,Law2020,Hiramatsu2022,Lanman2021,DSA,Bhardwaj2023b,Ryder2022,Mahony2018,Ibik2023,Michilli2022,Driessen2023,DeepSynopticArrayTeam2022}}. 

\end{figure}

In Figure~\ref{fig:sfr_vs_mstar}, we illustrate a comparison between our golden sample galaxies and some well-known FRB hosts, which belong to different types with varying star formation histories, focusing on their star formation rate (SFR) and stellar mass.
The color in the plot represents the redshift, while the dotted lines denote the boundary that distinguishes between star-forming and quiescent galaxies at various redshifts \citep{sfr_vs_mstar}.
It is evident that the majority of FRB hosts, as well as UGC\,10667 and SDSS\,J171046.84+212732.9, are characterized as star-forming galaxies. Their SFR predominantly falls within the range of $0.1$ to $10.0 \, {\rm M_{\odot}/yr}$.
Apart from this, we did not find any significant clustering or correlations among these FRB hosts. Therefore, similar to most of the literature, we also consider that FRBs may have multiple origins, and the current FRB host sample size is not large enough to distinguish them. Currently, or in the near future, there are powerful telescopes capable of accurately pinpointing the host galaxies of FRBs, such as DSA \citep{DSA} and BURSTT \citep{BURSTT}, or ASKAP \citep{ASKAP_new}and CHIME \citep{CHIME/FRBcatlog12021} in combination with outriggers \citep{VLBI}. However, because of the difficulty in obtaining spectral information from galaxies, the number of useful host galaxy samples remains limited. In the future, more powerful telescopes are expected to gather more information on host galaxies, improving our understanding of the origins and classifications of FRBs.

In this study, we conducted another comparison by comprehensively examining these two galaxies alongside NGC4993, i.e. the host galaxy of the KN AT\,2017gfo. NGC4993 is a galaxy with a prominent bulge and a mean stellar mass of $(0.3-1.2)\times 10^{11}\,{\rm M_{\odot}}$. Analysis of its spectral energy distribution indicates negligible star formation activity \citep{Im2017}. Despite variations in stellar mass and SFR between UGC\,10667, SDSS\,J171046.84+212732.9 and NGC4993, these three galaxies exhibit similar mean stellar ages of $\rm 2 \sim 3\, Gyr$ \citep{Im2017}, and metallicities spanning from 20\% to 100\% of the solar abundance \citep{Im2017}. These common characteristics align them with some host galaxies of short GRBs \citep{Graham2013,Contini2018,Yu2022}, whose origins are BNS mergers. Therefore, this result tends to support the hypothesis that UGC\,10667 or SDSS\,J171046.84+212732.9 was a possible host galaxy that could have produced a BNS merger, generating GW190425 and subsequently FRB\,20190425A.

\begin{figure}
\begin{center}
  \includegraphics[width=0.95\columnwidth]{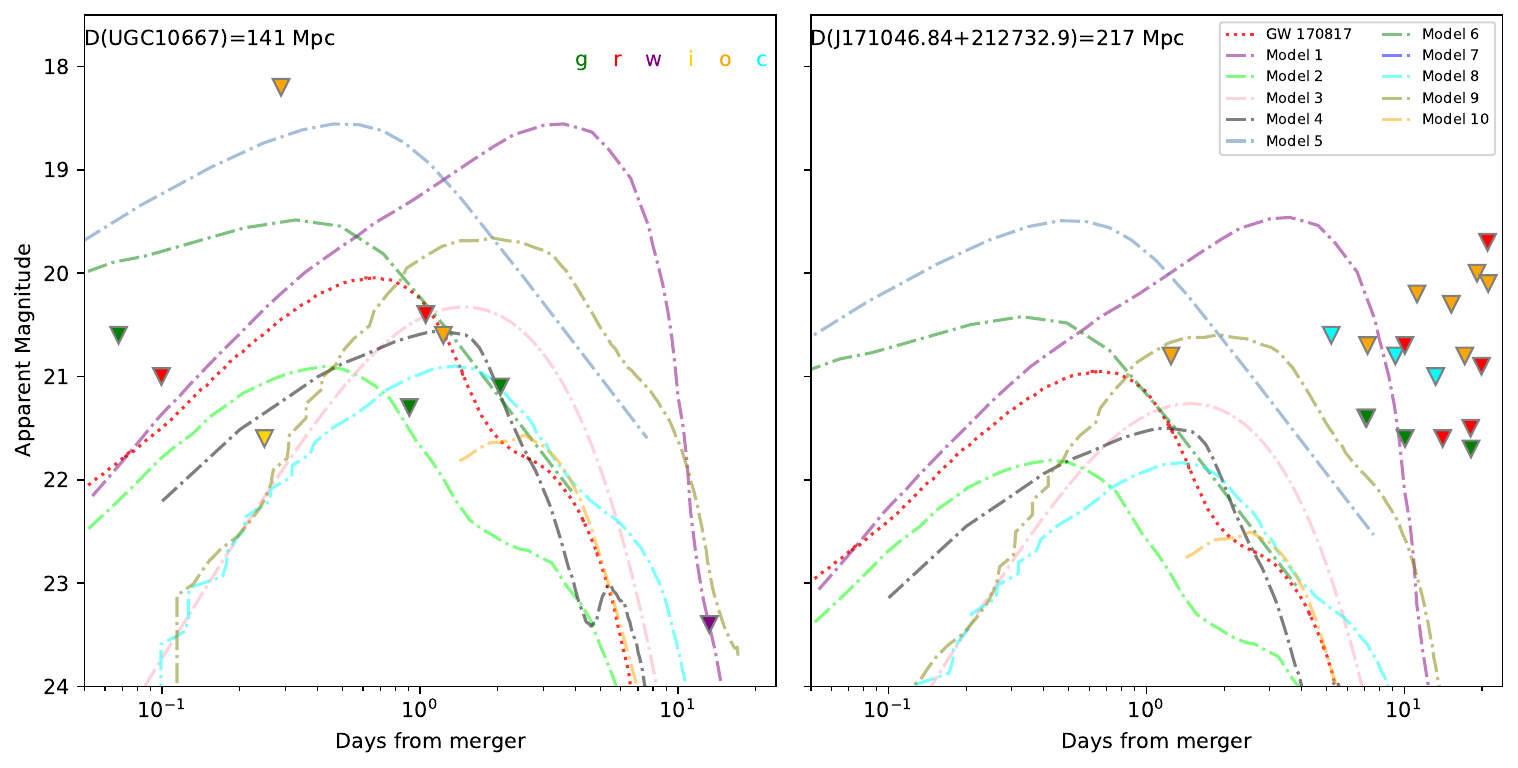}\\
  \end{center}
  \caption{\textbf{The comparison between Kilonova models and observed optical emission of UGC\,10667 and SDSS\,J171046.82+212732.9.} In both panels, colored curves represent kilonova light curve models described in Sec \ref{subsec:kncons}, all calculated in the r-band and assuming distances to UGC\,10667 or SDSS\,J171046.82+212732.9. The inverted triangles in the left panel denote the $3.5\sigma$ limits of the Pan-STARRS1, ATLAS, and ZTF images for UGC\,10667 around the time of GW190425, while those in the right panel represent the $3.5\sigma$ limits of the ATLAS and ZTF images for SDSS\,J171046.82+212732.9 during the same period.}
\label{fig:knmodel}
\end{figure}

\begin{table}
\renewcommand{\arraystretch}{1.2}  
\setlength{\tabcolsep}{30pt}
\begin{center}
    \begin{tabular}{ccccc}\hline\hline
$\rm MJD$ & $\rm Epoch$ & $\rm Telescope$ & $\rm Filter$ & $\rm limit$\\ \hline
$ 58599.590000$ & $ +1.24411$ & $\rm ATLAS$ & $o$ & $>20.8$\\
$ 58603.550000$ & $ +5.20411$ & $\rm ATLAS$ & $c$ & $>20.6$\\
$ 58605.420185$ & $ +7.07430$ & $\rm ZTF$ & $r$ & $>21.4$\\
$ 58605.483345$ & $ +7.13745$ & $\rm ZTF$ & $g$ & $>21.4$\\
$ 58605.540000$ & $ +7.19411$ & $\rm ATLAS$ & $o$ & $>20.7$\\
$ 58607.570000$ & $ +9.22411$ & $\rm ATLAS$ & $c$ & $>20.8$\\
$ 58608.380023$ & $ +10.03413$ & $\rm ZTF$ & $r$ & $>20.7$\\
$ 58608.402720$ & $ +10.05683$ & $\rm ZTF$ & $g$ & $>21.6$\\
$ 58609.530000$ & $ +11.18411$ & $\rm ATLAS$ & $o$ & $>20.2$\\
$ 58611.560000$ & $ +13.21411$ & $\rm ATLAS$ & $c$ & $>21.0$\\
$ 58612.420197$ & $ +14.07431$ & $\rm ZTF$ & $r$ & $>21.6$\\
$ 58613.540000$ & $ +15.19411$ & $\rm ATLAS$ & $o$ & $>20.3$\\
$ 58615.480000$ & $ +17.13411$ & $\rm ATLAS$ & $o$ & $>20.8$\\
$ 58616.397072$ & $ +18.05118$ & $\rm ZTF$ & $r$ & $>21.5$\\
$ 58616.461250$ & $ +18.11536$ & $\rm ZTF$ & $g$ & $>21.7$\\
$ 58617.460000$ & $ +19.11411$ & $\rm ATLAS$ & $o$ & $>20.0$\\
$ 58618.241771$ & $ +19.89588$ & $\rm ZTF$ & $r$ & $>20.9$\\
$ 58619.353113$ & $ +21.00722$ & $\rm ZTF$ & $r$ & $>19.7$\\
$ 58619.470000$ & $ +21.12411$ & $\rm ATLAS$ & $o$ & $>20.1$\\
$ 58627.377894$ & $ +29.03200$ & $\rm ZTF$ & $r$ & $>21.4$\\
$ 58627.410000$ & $ +29.06411$ & $\rm ATLAS$ & $o$ & $>20.6$\\
    \hline
    \end{tabular}
\end{center}
\vspace{-2mm}  
\caption{\label{tab:optlim} \textbf{The $\mathbf{3.5\sigma}$ limits of the ATLAS and ZTF images of SDSS\,J171046.82+212732.9 around the time of GW190425.} The epoch represents the time when the image was captured relative to the merger of GW190425.}
\end{table}

\section{Further Analyze Assuming GW/FRB Association\label{sec:fa}}
Though it cannot now be proven, if in fact GW190425 and FRB 20190425A are associated, we use this as an example to determine the most likely host galaxy by searching for KN emission. We further constrain the Hubble constant by combining the redshift information of the host galaxy with the distance estimate from GW190425.

\subsection{Constraints on Kilonova Emission\label{subsec:kncons}}
To explore the possibilities of these two galaxies hosting KN events, we constrained their reported observation limits to various KN models as well as compared them to the KN AT\,2017gfo.
As shown in Figure~\ref{fig:knmodel}, the inverted triangles in the left panel of Figure~\ref{fig:knmodel} represent the $3.5\sigma$ limits of the Pan-STARRS1, ATLAS and ZTF images for UGC\,10667 around the time of GW190425 \citep[as listed in the Table 1 of ][]{Smartt2024}, while those in the right panel represent the $3.5\sigma$ limits of the ATLAS and ZTF images for SDSS\,J171046.82+212732.9. Those limits are extracted from the ATLAS \citep{atlas2018,atlasfp} and ZTF forced photometry services \citep{ztffp} through the $\rm \textbf{HAFFET}$ program \citep{haffet} (as listed in Talbe\,\ref{tab:optlim}).
KN exhibits a pronounced blueness in the early stage, which transitions to deep red shortly thereafter. In the optical band, most of the KN emission is concentrated around the r-band, with observations in other bands being similar in brightness or fainter than those in the r-band. Hence, for simplicity, we exclusively present the results for the r-band in Figure\,\ref{fig:knmodel}.
For comparison, we overplotted the Lightcurve of KN AT\,2017gfo with best-fit parameters from \cite{Nicholl2021}, with the distance adjusted accordingly.

Various KN models are illustrated as dashed lines.
Model 1 is a magnetar-enhanced KN emission model with the same parameters as in \cite{Smartt2024}.
Model 2 is the same as AT\,2017gfo with a maximum stable NS mass $\rm M_{TOV} >2.63\, M_{\odot} $.
For Models 3 and 4, the assumption is a radioactive-powered emission with an ejecta mass of $\rm 10^{-2} \, M_{\odot}$, an outflow speed $v=0.1c$, iron-like opacities, and a thermalization efficiency of 1 (for Model 4) and blackbody emission (for Model 3 \citep{Lira1998}) with the same values of mass and velocity \citep{Metzger2010}.
Models 5 and 6 are kilonova models of \cite{Piran2013}. The approximation is that all bolometric luminosity is concentrated in the R band, assuming a BNS merger with $\rm M_{NS} = 1.4\, M_{\odot}$ and a black hole (BH)-NS merger with $\rm M_{NS} = 1.4\, M_{\odot}$ and BH mass of $\rm 10\, M_{\odot}$. This assumes low velocity ($0.1c$) and low ejecta mass ($\rm 10^{-3}\, M_{\odot}$) \citep{Barnes2013}.
Model 7 assumes an ejected mass of $10^{-3}$ M$_{\odot}$, a velocity of $0.1c$ and lanthanide opacity \citep{Barnes2013}.
Models 8 and 9 assume an accretion disc mass of $\rm 0.03 \, M_{\odot}$ and a remnant hypermassive NS or a remnant NS collapsing into a BH within 100 ms \citep{Kasen2015}.
Model 10 represents a BH-NS merger with a BH/NS mass ratio of 3, ejected mass of $0.0256\, \rm M_{\odot}$, velocity $v=0.237c$, a hard equation of state for the NS and BH spin of 0.75 \citep{Kawaguchi2016}.
All models are scaled according to the distances of UGC\,10667, and SDSS\,J171046.82+212732.9, respectively.

While the merger of a BNS system generates a GW signal, it may also give rise to short GRBs and KN. However, no associated GRB and KN were detected by following observations of GW190425 and FRB\,20190425A.
The absence of prompt emission from a short GRB could be due to the directional nature of its jet, where the narrow opening angle is not aligned with Earth, rendering it unobservable. In contrast, KN emissions are isotropic. So, in principle, we should detect them only if the observations reach deep enough in time.
\cite{Smartt2024} explored the KN emission for UGC\,10667 with respect to the BNS merger time window referred to in GW190425, and found no plausible KN emissions. This work used a corresponding method applied to SDSS\,J171046.84+212732.9 with similar results.
Although KN emissions were not ultimately detected, we can still explore whether these galaxies (i.e., UGC\,10667 and SDSS\,J171046.84+212732.9) might have hosted faint KN events, by scaling the KN AT\,2017gfo and various KN models to the distance of the galaxy and comparing them to detection limits.
We investigated this in Figure~\ref{fig:knmodel}, using the types of KN models mentioned above. Here, we provide a concise summary of our findings:
1) Similar to the study on UGC\,10667 conducted in \cite{Smartt2024}, we found that for SDSS\,J171046.84+212732.9, KN models involving magnetar-enhanced kilonova emission are ruled out by the optical upper limits, particularly as they slowly approach their peak, i.e., more than ten days.
2) Even though SDSS\,J171046.84+212732.9 is farther away, which could better explain the absence of a KN, UGC\,10667 still satisfies the majority of KN models. Therefore, the results of KN searching do not provide conclusive evidence to determine which of them is the true host of GW190425 and FRB\,20190425A.
3) The detection range of LIGO for BNS mergers is predicted to be more than 100 Mpc \citep{lvk2020}. At such distances, KN would be fainter than most of the current optical surveys based on their current schedules. In the future, with the advent of more powerful telescopes, such as the Rubin Observatory's Legacy Survey of Space and Time (LSST) \citep{lsst2019}, much fainter optical limits can be achieved. This will aid in detecting KNe as well as pinpointing the most likely host galaxy.

\begin{figure}
\begin{center}
  \includegraphics[width=130mm]{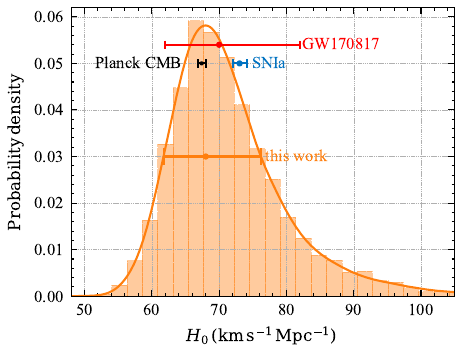}\\

\end{center}
\caption{\label{fig:h0}\textbf{The hyper posterior distribution using 95.5\% $P_{tmd}$ events.} The red point and errorbar indicate the result of GW170817 with $H_0=70.0^{+12.0}_{-8.0}$~\hunit. 
The black (blue) point and errorbar correspond to the measurements obtained from CMB (SN Ia) with $H_{0}=67.4 \pm 0.5\,(73.04 \pm 1.04)$~\hunit. 
The orange lines show the $H_0=68.0_{-6.2}^{+8.2}$~\hunit based on the combination GW190425, FRB\,20190425A and the $H_0$ posterior of GW170817. All results are at $1\sigma$ uncertainty.
}
\end{figure}

\begin{figure}
\begin{center}
  
  \includegraphics[width=0.9\columnwidth]{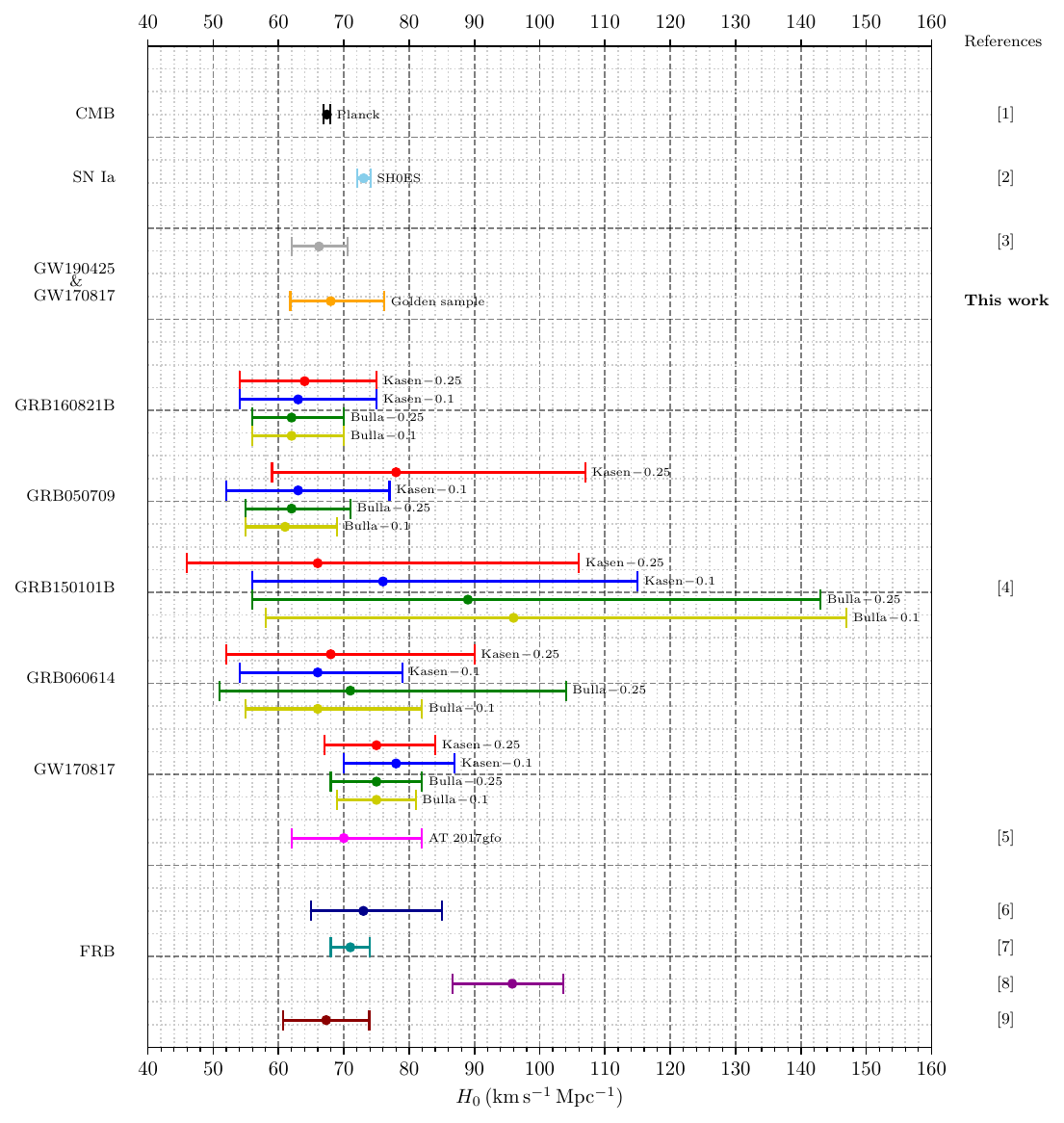}\\
  \end{center}
  \caption{\textbf{Hubble constant determined by different measurement or kilonova model.} 
  The points and errorbars in black and sky-blue represent the results from Planck CMB and SH0ES SN Ia, respectively. The gray point and errorbar denote the combination result of GW170817 and GW190425. 
  The orange point and errorbar represent the results of the golden sample in our study, while the red, blue, green, and dark yellow points and errorbars denote the results from light curves associated with short GRBs, assuming that these are attributable to KN with different models.
  The magenta point and error bar represent the results of AT\,2017gfo and GW170817.
  The points and error bars in dark blue, dark cyan, dark magenta, and dark red represent the results obtained using different methods based on the FRB data.
  All errorbars in the results are presented with an uncertainty of $1\sigma$. 
  \protect\\
  References:[1] \cite{Planck2018}; [2] \cite{sh0es}; [3] \cite{Dietrich2020sci}; [4] \cite{Coughlin2020}; [5] \cite{Abbott2017GW170817H0}; [6] \cite{James2022}; [7] \cite{Liu2023}; [8] \cite{Wei2023}; [9] \cite{Fortunato2024}.}
  
\label{fig:h0compare}

\end{figure}

\subsection{Hubble Constant Estimation\label{subsec:h0est}}
Given the fact that current observations cannot definitively identify the real host galaxy, we therefore opt to constrain the Hubble constant using both SDSS\,J171046.84+212732.9 and UGC\,10667. Moreover, for a more comprehensive consideration, a similar procedure is applied to the golden sample galaxies, which cover most of the significance corresponding to the generation of GW190425 and FRB\,20190425A. The Bayesian framework for estimating $H_0$ is shown below.

The $p(d_{\rm L},z^i | d^i)$ is denoted as the posterior distribution of luminosity distance ($d_{\rm L}$) and redshift for the GW190425 and the possible $i$-th host galaxies of FRB\,20190425A, based on observations dataset $d^i$. 
Our primary objective is to identify an effective model capable of describing the set of luminosity distance and redshift posteriors for $N$ host galaxies. 
Here introduce a conditional prior $\pi(d_{\rm L},z^{i} | \Lambda)$ for luminosity distance from gravitational wave detection and redshifts of host galaxies, incorporating the hyperparameters  $\Lambda$.
The hyperposterior distribution of $\Lambda$ is defined by:

\begin{equation}
\label{eqn:hyperpost}
p(\Lambda|{\bf{d} } ) 
= \frac{1}{{\cal Z}_{\Lambda } } 
\sum^{N}_{i=1} \int {\rm d}d_{\rm L} {\rm d}z^i P^{i}_{tmd} {\cal L}(d^i| d_{\rm L}, z^i) \pi( d_{\rm L}, z^i|\Lambda) \pi(\Lambda) \frac{1}{{\cal{N}}_{s}(\Lambda)}\ .
\end{equation}
Where $\pi (\Lambda )$ represents the prior distribution for hyperparameters $\Lambda$, while the term ${\cal L}(d^i| d_{\rm L}, z^i)$ denotes the likelihood function of the data given $d_{\rm L}$ and $z^i$. 
The hyperparameters $\Lambda$ are the Hubble constant $H_0$ and the matter density $\Omega_m$. 
${\cal{N}}_{s}(\Lambda)$ is the selection effect term, described as ${\cal{N}}_{s}(\Lambda) \propto H_0^3$\cite{Abbott2017GW170817H0,Dietrich2020sci}.
The ${\bf d}$, constructed by $d^i$, refers to the original observations providing measurements of luminosity distance and redshift. The likelihood related to the luminosity distance and redshift posteriors with original observations $d^i$ is described by \cite{Eric2018}:
\begin{equation}\label{eqn:mlikelihood1}
{\cal L}(d^i| d_{\rm L}, z^i) = \frac{{\cal Z}_0^i}{\pi_0(d_{\rm L}, z^i)} p(d_{\rm L}, z^i|d^i)\ .
\end{equation}
Here, $\pi_0(d_{\rm L}, z^{i})$ is the flat initial prior, a constant, used to derive the luminosity and redshift posterior, and ${\cal Z}_0^i$ represents the initial evidence that cancels out during hyperposterior normalization or Bayesian evidence ratio constructions. The simplified likelihood is written as:
\begin{equation}\label{eqn:mlikelihood2}
{\cal L}(d^i| d_{\rm L}, z^i) \sim  p(d_{\rm L}, z^i|d^i)\ .
\end{equation}
The  hyperevidence ${\cal Z}_\Lambda $ term in Equation (\ref{eqn:hyperpost}) is formulated as:
\begin{equation}
\label{eq:ZLambda}
{\cal Z}_{\Lambda}  = \int {\rm d}\Lambda 
\sum ^N_{i=1} \int {\rm d}d_{\rm L} {\rm d}z^i P_{tmd}^i {\cal L}(d^i|d_{\rm L}, z^i)
\pi(d_{\rm L}, z^i|\Lambda) \pi(\Lambda)\frac{1}{{\cal{N}}_{s}(\Lambda)}.
\end{equation}

To practically compute this, we replace the integral over ${\rm d} d_{\rm L}$ and ${\rm d}z^i$ with a summation using posterior samples of $d_{\rm L}$ and $z^i$, see \cite{Eric2018}. This approach leads to a more tractable expression for the hyperevidence based on the flat initial prior and the canceled initial evidence:
\begin{equation}
\label{eq:ZLambda_samp1}
{\cal Z}_{\Lambda}  \sim \int {\rm d}\Lambda 
\sum _{i=1}^N  P^{i}_{tmd} \sum^{n_i}_{k=1} \pi(d_{\rm L},z^i_k|\Lambda) \pi(\Lambda)\frac{1}{{\cal{N}}_{s}(\Lambda)}, 
\end{equation}
as well as the hyperlikelihood, 
\begin{equation}
\label{eq:ZLambda_samp2}
{\cal L}_{\Lambda}  \sim 
\sum _{i=1}^N P^{i}_{tmd} \sum^{n_i}_{k=1} \pi(d_{{\rm L}\,,k},z^i_k|\Lambda).
\end{equation}
Where $n_i$ refers to the number of redshift posterior samples (equivalent to those of luminosity distance, employing $n_i=10^4$ in this work) for the $i$-th possible host galaxies.

In this work, we adopt a flat Friedmann-Robertson-Walker Universe and consistently employ the $\Lambda$CDM model as a reference throughout this study. The expression for the luminosity distance is given by:
\begin{equation}
    d_{\rm L}(z,H_0,\Omega_m)=(1+z){\int^z_0}c/H(z')dz'\, ,
\end{equation}
where the Hubble parameter, $H(z)$, assumes a dark energy equation of state of $w=-1$. This parameter is derived as $H(z)=H_0\sqrt{(1+z)^3\Omega_m+\Omega_{\Lambda}}$, with $\Omega_m$ representing the matter density and $\Omega_{\Lambda}=1-\Omega_m$ denoting the dark energy density. The comoving distance, $r(z)$, is calculated as $r(z) = d_{\rm L}/(1+z)$. Under the assumption that the luminosity distance obtained by gravitational wave observation, GW190425, is associated with redshifts from host galaxy candidates of FRB\,20190425A, the conditional prior $\pi(d_{\rm L}, z^i| \Lambda)$ can be expressed as,
\begin{equation}
\pi(d_{\rm L}, z^i | H_0, \Omega_m, \sigma)=\frac{1}{\sigma \sqrt{2 \pi}} \exp \left(-\frac{[d_{\rm L}- d_{\rm L}(z^i, H_0,\Omega_m) ]^2}{2 \sigma^2}\right) \text {. }
\end{equation}

Utilizing $P_{tmd}$ as weight and incorporating the redshift of the galaxy, the combination of the luminosity distance posterior of GW190425 and the $H_0$ posterior of GW170817 yields an estimate for the Hubble constant as $H_0=68.0_{-6.2}^{+8.2}\ {\rm km\,s^{-1}\,Mpc^{-1}}$ for golden sample \footnote{The Hubble constant calculated with only SDSS\,J171046.84+212732.9 and UGC\,10667 closely resemble that of the golden sample, given their substantial contributions to the overall weight of $P_{tmd}$. Therefore, we discuss only the result of the golden sample throughout this work.}.
As shown in Figure~\ref{fig:h0}, by incorporating the contributions from GW190425 and its associated observations, we achieved a notable 30\% reduction in the $1\sigma$ uncertainty of $H_0$ compared to that of GW170817.
If the posterior uncertainty of the luminosity distance of GW is further reduced, we anticipate even more refined outcomes.
Given the large uncertainties in our results, it is difficult for them to offer much insight into arbitrating the Hubble tension. While our central value leans slightly toward the CMB result, both the CMB and SN Ia results lie well within our $1\sigma$ range.
In Figure~\ref{fig:h0compare}, we provide a comparative analysis of the results obtained in this study with other measurements, such as those from GW170817, CMB, SN Ia and FRB,  as well as some specific BNS mergers with both short GRB observation and probable KN emissions. As shown, most of the results, including ours, have a mean value between $60$ and $80\,$\hunit. 

\section{Conclusion}
In this paper, we introduce a novel approach to identifying potential host galaxies, using the assumed association of GW190425 and FRB\,20190425A as an example to demonstrate our methodology.
With the CHIME localization map, we re-search the host galaxy of FRB\,20190425A and found another galaxy, SDSS\,J171046.84+212732.9, with a significant $P_{tmd}$ and an even higher $P_{\rm \, PATH}$ compared to UGC\,10667.
From the perspectives of $P_{tmd}$ and $P_{\rm \, PATH}$, we cannot distinguish between these two galaxies well. Therefore, we conducted further exploration, such as fitting and comparing them with known FRB host galaxies, and fitting KN models to their reported observation limits. With these attempts, we found that these two galaxies are very similar, e.g. they had similar age, metalicity, and star formation history. Although SDSS\,J171046.84+212732.9 is farther away which can better explain the absence of any detected KN emission, UGC\,10667 can also satisfy the majority of KN models.
Hence, based on the existing observations, we are unable to ascertain which galaxy serves as the actual host galaxy.
Consequently, we used both of them (and galaxies with a cumulative $P_{tmd}$ at 95\% after normalization, taking a more comprehensive perspective into account) with their $P_{tmd}$ as a weight to constrain the Hubble constant.
Along with the $H_0$ posterior from GW170817, the combined constraints yield an improved result for $H_0$ at $1\sigma$ uncertainty, with $H_0=68.0^{+8.2}_{-6.2} \,{\rm km\,s^{-1}\, Mpc^{-1}}$.
Although the uncertainty in our results prevents us from arbitrating the tension between SN Ia measurements and the CMB, the inclusion of the redshift of the host galaxy identified through the FRB localization tends to push toward a lower value for $H_0$.

Measurements from CMB and SN Ia are more precise, as they are well-observed; however, they do not fall within the same range. The Hubble tension, often attributed to errors, assumptions, or potential new physics, remains a topic of interest. The inclusion of GW observations, especially those with well-localized events, presents a new avenue to explore and refine Hubble constant constraints. In this study, we demonstrate that well-localized GW events (with efforts of FRB localization) can contribute to a more accurate determination of $H_0$, despite lacking plausible EM counterpart measurements.
We anticipate that in the future there will be more well-localized GW events, especially those associated with GRB/FRB/KN. The joint constraints from these events on the Hubble constant are expected to improve in precision significantly. Using GW as a standard siren for cosmological constraints will likely yield more achievements as LIGO-Virgo-KAGRA continues its operations. This includes potential contributions to arbitrating the Hubble tension.

\begin{acknowledgments}
\nolinenumbers

The authors thank anonymous referees for their thoughtful comments and feedback, which have strengthened this manuscript. SY acknowledges the funding from the National Natural Science Foundation of China under Grant No. 12303046. ZQY is supported by the National Natural Science Foundation of China under Grant No. 12305059. The ZTF forced-photometry service was funded under the Heising Simons Foundation grant No. 12540303 (PI: Graham). 
\end{acknowledgments}

\bibliography{ref}{}
\bibliographystyle{aasjournal}

\end{document}